\documentclass[superscriptaddress,notitlepage,nofootinbib]{revtex4-1}
\usepackage{natbib}
\usepackage{physics}            
\usepackage{amsmath, amsfonts}
\usepackage{siunitx}
\usepackage{graphicx}
\usepackage{xcolor}

\usepackage[english]{babel}
\usepackage[utf8]{inputenc}
\usepackage[autostyle]{csquotes}
\MakeOuterQuote{"}
\usepackage[shortlabels]{enumitem}
\usepackage{dsfont}
\usepackage[hidelinks]{hyperref}	
\usepackage[capitalize]{cleveref}

\usepackage{pgf}
\usepackage{stackengine}

\newcommand{\blue}[1]{{#1}}

\begin{document}
	
	\title{Twin-field quantum key distribution with fully discrete phase randomization}
	
	\author{Guillermo \surname{Currás Lorenzo}}
	\email{g.j.curraslorenzo@leeds.ac.uk}
	\affiliation{School of Electronic and Electrical Engineering, University of Leeds, Leeds, UK}
	
	\author{Lewis \surname{Wooltorton}}
	\affiliation{School of Electronic and Electrical Engineering, University of Leeds, Leeds, UK}
	
	\author{Mohsen \surname{Razavi}}
	\affiliation{School of Electronic and Electrical Engineering, University of Leeds, Leeds, UK}

	\begin{abstract}
		Twin-field (TF) quantum key distribution (QKD) can overcome fundamental secret-key-rate bounds on point-to-point QKD links, allowing us to reach longer distances than ever before. Since its introduction, several TF-QKD variants have been proposed, and some of them have already been implemented experimentally. Most of them assume that the users can emit weak coherent pulses with a continuous random phase. In practice, this assumption is often not satisfied, which could open up security loopholes in their implementations. To close this loophole, we propose and prove the security of a TF-QKD variant that relies exclusively on discrete phase randomisation. Remarkably, our results show that it can also provide higher secret-key rates than counterpart protocols that rely on continuous phase randomisation.

	\end{abstract}

	\maketitle

	\section{Introduction}
	
	Quantum key distribution (QKD) allows two users, Alice and Bob, to generate a shared secret key in the presence of an eavesdropper, Eve, with unlimited computational power. Despite its great potential, QKD has yet to overcome important practical problems before it is ready for widespread use. One of the most important challenges is how to perform QKD at long distances, given that, in optical fibres, the loss increases exponentially with the channel length. Even with a GHz repetition rate, it would take 300 years to successfully send a single photon over 1000 km of standard optical fibres \cite{sangouard2011quantum}. Another crucial issue is to guarantee that a particular implementation of a QKD protocol is secure. That is, we have to either show that our implementation satisfies all assumptions made in the corresponding theoretical security proof, or to devise a security proof that matches the realities of an experiment. In this work, we do the latter for one of the most advanced QKD protocols, known as twin-field QKD (TF-QKD) \cite{lucamarini2018overcoming}, as one of the key candidates for improving key rate scaling with distance.
	
	Fundamental bounds show that the key rate of repeaterless QKD protocols scales at best linearly with $\eta$ \cite{pirandola2017fundamental}, where $\eta$ is the transmittance of the channel connecting Alice and Bob. TF-QKD breaks this limitation by offering a key rate that scales with $\sqrt{\eta}$.  The key enabling idea behind TF-QKD operation is to effectively generate an entangled state between the two users in the space spanned by vacuum and single-photon states. To do so, we need a repeater node that performs entanglement swapping, using single-photon interference, as well as phase stability across the channel to make sure the generated state is in the desired superposition form. This approach requires only one photon to survive the path loss over half of the channel, thus the improved scaling with distance. Note that TF-QKD is not the only protocol that achieves this scaling. There are other protocols, inspired by quantum repeater structures, that achieve the same rate scaling by using quantum memories \cite{panayi2014memory,abruzzo2014measurement} and/or quantum non-demolition measurements \cite{azuma2015all}. But, TF-QKD is, experimentally, in a more advanced state than such alternatives. In fact, certain variants of TF-QKD have already been implemented experimentally \cite{minder2019experimental, zhong2019proof, liu2019experimental, wang2019beating}, and a distance record exceeding 500~km has already been achieved \cite{chen2020sending}. This would make the implementation security issue crucially relevant for TF-QKD experiments.
	

	One of the key practical constraints on a QKD system is imposed by the type of optical source/encoder needed in the respective protocol. The corresponding security proof would then need to match such practical constraints. The single-photon version of TF-QKD has a simple theoretical description \cite{curty2019simple}, but it is difficult to implement in practice. Thus, a significant research effort has focused on developing practical variants \cite{curty2019simple,wang2018twin,ma2018phase,cui2019twin} in which the users encode weak coherent pulses (WCPs). These variants differ in their protocol descriptions and/or security proofs, but, so far, all of them rely on the decoy-state method \cite{lo2005decoy}. That is, they either use decoy states in their key mode \cite{wang2018twin,ma2018phase}, i.e., to generate the key, and/or in their test mode \cite{curty2019simple,cui2019twin}, i.e., to estimate Eve's side information on the key. 
	
	
	
	Conventional decoy-state techniques require the use of phase-randomised coherent states (PRCS).	This implies that the users are ideally able to randomise the global phase of their pulses \textit{continuously} and \textit{uniformly}. This is, however, difficult to achieve in practice. Experimentally, there are two approaches to randomise the phase of a coherent pulse: passive and active. Passive randomisation consists of turning the laser off and then on again to generate the PRCS. In addition to the impracticality of this approach in a high-rate QKD system, it is hard to guarantee experimentally that the generated phase genuinely follows a uniform distribution \cite{cao2015discrete}. In fact, experiments have shown that, in practice, there are phase correlations between adjacent pulses \cite{xu2012ultrafast,abellan2014ultra}. In an active randomisation procedure, a phase modulator is used, in combination with a random number generator. This approach fits the TF-QKD variant of Refs.~\cite{curty2019simple,cui2019twin} very well, since one already needs a phase modulator to produce the phase-locked coherent states sent in the key mode. However, it randomises the phase over a \textit{discrete}, not \textit{continuous}, set of values. Thus, none of these two approaches necessarily satisfy the assumptions of the decoy-state method, which could open security loopholes in the experimental implementations of TF-QKD.
	
	In this work, we address such security loopholes, in the context of TF-QKD, by proposing a variant that relies on discrete phase randomisation in both key and test modes. The quantum phase of our protocol is similar to that of Refs.~\cite{curty2019simple,cui2019twin}, with the main difference being that we use discrete, not continuous, phase randomisation in the test mode as well. This would allow us to use additional post-selection in the test mode when users have both used \textit{matching} phase values, i.e., either exactly the same phases, or those with a phase difference of exactly $\pi$. The concept of phase discretisation, and its corresponding post-selection, has also appeared in some other TF-QKD variants. For instance, in phase-matching QKD \cite{ma2018phase}, phase slices are introduced for post-selection. Nevertheless, they still rely on continuous phase randomisation for their encoded states. The same applies to the test mode of the protocol introduced in \cite{wang2019twinfield}, in which they generalise the key-mode encoding space in \cite{curty2019simple,cui2019twin} from a two-state discrete space to a multiple-state discrete space. In sending-or-not-sending variant, in both modes, we rely on continuous phase randomisation. By totally removing the continuous phase randomisation, in this work, we then offer a security proof that is aligned with the realities of the existing implementations. Table~\ref{tab:DP-comp} shows how different TF-QKD variants are compared in terms of using discrete phase randomisation.    
	
	\begin{table}[h]
	    \centering
	    \begin{tabular}{c|c|c|c}
	    \hline
	         Protocol & Key mode & Test mode & Phase-based Post-selection \\ \hline
	          Phase Matching QKD \cite{ma2018phase} & Continuous  & Continuous & Yes (both modes)\\ \hline
	         Sending-or-not-sending TF-QKD \cite{wang2018twin} & Continuous  & Continuous & Yes (test mode)\\ \hline
	         TF-QKD in Refs.~\cite{curty2019simple,cui2019twin} & Discrete  & Continuous & No\\ \hline
	         Discrete Phase Matching QKD \cite{wang2019twinfield} & Discrete  & Continuous & Yes (test mode)\\ \hline  
	         This work & Discrete & Discrete & Yes (test mode) \\ \hline
	    \end{tabular}
	    \caption{A comparison between major TF-QKD variants and whether they use continuous or discrete phase randomisation for their key/test modes. The post-selection column specifies if in the post-processing stage, signals with similar phases will be post-selected or not.}
	    \label{tab:DP-comp}
	\end{table}
	
	One of the key contributions of this work is to offer a rigorous security proof for TF-QKD with discrete phase randomisation in its test mode. Note that discrete phase randomisation has already been considered in Ref.~\cite{cao2015discrete} for a decoy-state BB84 protocol. In \cite{cao2015discrete}, discrete phase randomisation can be considered as a source flaw. Consequently, they find that, while the secret key rate obtainable using discrete randomisation is always strictly worse than using continuous randomisation, the former quickly approaches the latter as the number of discrete random phases increases. In fact, one can obtain a performance reasonably close to the continuous case using as few as ten random phases. However, it is not immediately clear whether this behaviour will hold for the TF-QKD variants in \cite{curty2019simple,wang2018twin,ma2018phase,cui2019twin}, given that: (i) their security proofs are quite diverse, and some of them are very different from that of decoy-state BB84; and (ii) in TF-QKD, both users emit quantum states, thus the source flaw is present in both users. In fact, recent works have found that the security issue arising from flawed sources that leak information has a much bigger impact in measurement-device-independent (MDI) QKD \cite{wang2020measurement} than in BB84 \cite{wang2018finite}. In principle, the same could be true for other kinds of source imperfections, such as the use of discrete phase randomisation.

	Surprisingly, in our protocol, we realise that, by using discrete phase randomisation, the key rate can actually improve over the TF-QKD variant in Ref.~\cite{curty2019simple}. 	This is because discrete phase randomisation allows the users to postselect the detected test mode rounds in which their encoded phase values perfectly match. This postselected data turns out to result in a tighter estimation of the phase-error rate compared with the approach in Ref.~\cite{curty2019simple}, which relies on continuous phase randomisation without phase postselection. The fact that, in TF-QKD, users may share the same phase reference would also make this post-selection process practically possible---something which may not be available in BB84 or MDI-QKD protocols.

	Similar to some other protocols that rely on discrete randomisation \cite{cao2015discrete}, we use numerical techniques as part of our security proof. In particular, we use semidefinite programming (SDP) techniques, inspired by the work of Ref.~\cite{primaatmaja2019versatile}, to estimate the phase-error rate. We note that, in Ref.~\cite{primaatmaja2019versatile}, the authors already apply their generic numerical technique to prove the security of a TF-QKD protocol with fully discrete phase randomisation. However, their procedure is only efficient for cases in which only a few discrete phases is used. Here, we exploit the particularities of our protocol to introduce an analysis that uses a much smaller number of carefully chosen constraints. The corresponding SDP can efficiently be solved even with a large number of discrete phases, which allows us to investigate how the key rate improves when increasing the number of phase values.

	
	\section{Methods}
	
	\subsection{Protocol description}
	
	Our protocol is very similar to that of Refs.~\cite{curty2019simple,cui2019twin}.  Alice and Bob send quantum signals to an untrusted middle node Charlie, who (ideally) interferes them at a balanced 50:50 beamsplitter, performs a photodetection measurement, and reports the outcome. These signals belong to one of two "modes", key and test, selected at random. Key mode emissions are used to generate the raw key, while test mode emissions are used to estimate Eve's side information on the raw key. In key mode, the users send phase-locked coherent states $\ket{\pm \sqrt{\mu}}$. In test mode, the users send phase-randomised coherent states of different intensities. Unlike in Refs.~\cite{curty2019simple,cui2019twin}, the phases of the test-mode states are randomised over a \textit{discrete} set, rather than a continuous range. The detailed protocol steps are as follows:
	
	\small
	\begin{enumerate}[label*=(\arabic*)]
		\item \textit{Preparation} 
		
		Alice (Bob) randomly choose the transmission mode, key or test, and
		
		\begin{enumerate}[label=(\arabic{enumi}.\arabic*), ref=\arabic{enumi}.\arabic*] 
			\item \label{} If she (he) chooses key mode, she (he) generates a random bit $b_A$ ($b_B$), prepares an optical pulse in the coherent state $\ket{(-1)^{b_A} \sqrt{\mu}}$ ($\ket{(-1)^{b_B} \sqrt{\mu}}$), and sends it to Charlie.
			
			\item If she (he) chooses test mode, she (he) selects a random intensity $\beta_a$ $(\beta_b) \in \blue{\{\beta_1,\ldots, \beta_{d-2}, \mu, \beta_v\}}$, where $\mu$ is the same intensity used in key mode, and $\beta_v = 0$ is a vacuum intensity. Then, she (he) selects a random phase $\theta_a ~(\theta_b) = \frac{2 \pi m}{M}$ with $m \in \{0,1,2,\ldots,M-1\}$, prepares the state $\ket{\sqrt{\beta_a} e^{i \theta_a }}$ ($\ket{\sqrt{\beta_b} e^{i \theta_b }}$), and sends it to Charlie.
		\end{enumerate}
		
		\item \textit{Detection} 
		
		An honest Charlie interferes Alice and Bob's signals at a 50:50 beamsplitter, followed by threshold detectors $D_c$ and $D_d$ placed at the output ports corresponding to constructive and destructive interference, respectively. A round is considered successful if exactly one detector clicks, and unsuccessful otherwise. After the measurement, Charlie reports whether or not the round was successful, and, if it was, he reports which specific detector clicked.  

		\item \textit{Sifting} 
		
		For all successful rounds, Alice and Bob disclose their choices of key mode or test mode, keeping only data from those in which they have used the same mode. Then,
		
		\begin{enumerate}[label=(\arabic{enumi}.\arabic*), ref=\arabic{enumi}.\arabic*] 
			\item They calculate the gain $p_{\rm succ}$ of their key mode rounds, and generate their sifted keys from the values of $b_A$ and $b_B$ corresponding to these rounds. Then, they publicly disclose a small random subset of their sifted keys. With this information, they estimate the fraction of the sifted key, $p_{\rm same \vert succ}$ ($p_{\rm diff \vert succ}$), that originated from emissions in which their phase choices agreed (disagreed). Bob then flips his sifted key bits corresponding to the rounds in which $D_d$ clicked. Based on that, Alice and Bob estimate the bit error rate $e_{\rm bit}$.
			\item For all possible values of $\beta$, Alice and Bob calculate the gains $\{Q_{\beta}\}$ of the test mode rounds in which they both used intensity $\beta$ and the same phase $\theta_a = \theta_b$. They also calculate the gains $\{Q_{\beta}^{-}\}$ of the rounds in which they both used intensity $\beta$ and opposite phases $\theta_a = \theta_b \pm \pi$.
		\end{enumerate}

		\item \textit{Parameter estimation}
		
		Alice and Bob use the values of $\{Q_{\beta}\}$ and $\{Q_{\beta}^{-}\}$ to estimate the amount of key information $I_{AE}$ that may have been leaked to an eavesdropper.

		\item \textit{Postprocessing} 
		
		Alice and Bob perform error correction and privacy amplification to obtain a secret key.

	\end{enumerate}
	\normalsize
	
	Since this is a discretely-modulated MDI-type protocol, in principle, one could directly use the numerical techniques of Ref.~\cite{primaatmaja2019versatile} to prove its security. However, the SDP in Ref.~\cite{primaatmaja2019versatile} requires one constraint, in the form of an inner product, for each combination of emitted states. The number of different states sent in this protocol can then make such an approach infeasible in practice. Namely, since Alice and Bob send $\left[(d-1)M+1\right]^2$ different combinations of states\footnote{To compute the number of states, note that the set of test-mode states contains the set of key-mode states, so one only needs to count the former. Also, when Alice or Bob choose the vacuum intensity, they send the same vacuum state, independently of their choice of random phase.}, one needs to solve the dual problem of an SDP with $\left[(d-1)M+1\right]^4$  inner-product constraints, plus the constraints related to the measurement results of the protocol. Thus, even for $M=4$ and $d=3$, the simplest case considered in the numerical results of this paper, one needs to solve a SDP with more than 6561 constraints. For $M=12$ and $d=3$, the number of constraints grows to more than 390625. 
	This can make the implementation of such techniques infeasible on conventional computers \cite{LectureSDP,gondzio2012interior}. 
	
	In the following, we provide a security analysis that requires to solve the dual problem of two SDPs with only $(d-1)(d-2)M+ 2d+M-1$ constraints each. That is, for the examples considered above, we have SDPs with 17 and 41 constraints, respectively, which can be quickly solved using any commercial off-the-shelf laptop.
	
	\subsection{Security analysis}
	
	In our security analysis, we consider the asymptotic scenario in which the users emit an infinite number of signals. Also, for simplicity, we assume collective attacks. We note that, in the asymptotic regime, security against collective attacks implies security against general attacks, thanks to results such as the postselection technique \cite{christandl2009postselection}.
	
	We consider the virtual protocol in which Alice replaces her key mode emissions by the generation of the state
	\begin{equation}
	\ket{\psi}_{Aa} = \frac{1}{\sqrt{2}} \left( \ket{0}_A \ket{\sqrt{\mu}}_a + \ket{1}_A\ket{-\sqrt{\mu}}_a \right),
	\end{equation}
	where $A$ is a virtual qubit ancilla that she keeps in her lab, and $a$ is the photonic system sent to Charlie; and Bob replaces them by a similarly defined $\ket{\psi}_{Bb}$. We assume that Eve controls not only the quantum channels, but also the untrusted middle node Charlie, and the announcements he makes. From a security standpoint, we can describe Eve's collective attack as a two-outcome\footnote{Note that Charlie not only reports whether or not a round was successful, but also whether he obtained constructive or destructive interference. However, the latter announcement only determines whether or not Bob applies a bit flip, which does not affect Eve's side information. Thus, from a security standpoint, we can assume that Eve's general measurement only has two outcomes.} general measurement $\{\hat{M}_{ab},\hat{M}_{ab}^f\}$ on the photonic systems $ab$, where $\hat{M}_{ab}$ ($\hat{M}^f_{ab}$) is the Kraus operator corresponding to a successful (unsuccessful) announcement.  Conditioned on a success, Alice and Bob obtain a state,
	\begin{equation}
	\ket{\Psi}_{AaBb} = \frac{\hat{M}_{ab} \ket{\psi}_{Aa}\ket{\psi}_{Bb}}{\sqrt{p_{\text{succ}}}},
	\end{equation}
	where $p_{\text{succ}} = \norm{\hat{M}_{ab} \ket{\psi}_{Aa}\ket{\psi}_{Bb}}^2$ is the probability that Eve will announce a key mode round as successful.
	
	In our virtual protocol, after Eve's announcements, Alice and Bob perform the joint measurement $\{\hat{O}_{\rm same},\hat{O}_{\rm diff}\}$, with $\hat{O}_{\rm same} = \ketbra{00}_{AB} + \ketbra{11}_{AB}$ and $\hat{O}_{\rm diff} = \ketbra{01}_{AB} + \ketbra{10}_{AB}$, on the ancillas corresponding to the successful rounds, learning whether they used the same or different phases\footnote{Note that, to generate the raw key, Alice and Bob project their ancillas $A$ and $B$ in $\{\ket{0},\ket{1}\}$. Since the joint measurement commutes with this projection, it is a valid virtual protocol step.}. Depending on the result, they will obtain one of the two post-measurement states
	\begin{gather}
	\label{eq:Psi-same}
	\ket{\Psi_{\text{same}}} =  \frac{ \ket{00}_{AB} \hat{M}_{ab} \ket{\sqrt{\mu}}_a\ket{\sqrt{\mu}}_b + \ket{11}_{AB} \hat{M}_{ab} \ket{-\sqrt{\mu}}_a\ket{-\sqrt{\mu}}_b}{2 \sqrt{p_{\rm succ,same}}}, \\
	\label{eq:Psi-diff}
	\ket{\Psi_{\text{diff}}} =  \frac{ \ket{01}_{AB} \hat{M}_{ab} \ket{\sqrt{\mu}}_a\ket{-\sqrt{\mu}}_b + \ket{10}_{AB} \hat{M}_{ab} \ket{-\sqrt{\mu}}_a\ket{\sqrt{\mu}}_b}{2\sqrt{p_{\rm succ,diff}}},
	\end{gather}
	where $p_{\rm succ,same} = p_{\rm succ} p_{\rm same \vert succ} $ ($p_{\rm succ,diff} = p_{\rm succ} p_{\rm diff \vert succ}$) is the probability that Alice and Bob use the same (different) phases in a key mode round \textit{and} Eve reports the round as successful. This allows us to define the quantities
	\begin{gather}
	\label{eq:ph-error-same}
	e_{\text{ph,same}} =  \norm{\,_{AB}{\braket{++}{\Psi_{\text{same}}}}}^2 + \norm{\,_{AB}{\braket{--}{\Psi_{\text{same}}}}}^2, \\
	e_{\text{ph,diff}} =  \norm{\,_{AB}{\braket{++}{\Psi_{\text{diff}}}}}^2 + \norm{\,_{AB}{\braket{--}{\Psi_{\text{diff}}}}}^2,
	\end{gather}
	where $e_{\text{ph,same}}$ ($e_{\text{ph,diff}}$) is the phase-error rate of the successful key mode rounds in which Alice and Bob used the same (different) phases. Eve's side information of the sifted key (per key bit) can now be bounded by
	\begin{equation}
	I_{AE} \le p_{\rm same \vert succ} h(e_{\text{ph,same}}) + p_{\rm diff \vert succ} h(e_{\text{ph,diff}}),
	\end{equation}
	\blue{where $h(x) = -x \log_2 x - (1-x) \log_2 (1-x)$ is the Shannon binary entropy function}. The secret key rate that Alice and Bob can distill is
	\begin{equation}
	R \ge p_{\rm succ} \left[1 - I_{AE} - f h(e_{\rm bit})\right],
	\end{equation}
	\blue{where $f$ is the error correction inefficiency}.
	
	The objective of our security analysis is to obtain upper bounds on $e_{\rm ph,same}$ and $e_{\rm ph,diff}$, based on the data obtained in the test rounds. The procedure is very similar for both terms; we will first explain $e_{\rm ph,same}$.
	
	\subsubsection{Estimation of \texorpdfstring{$e_{\rm ph,same}$}{ephsame}}
	\noindent
	First, we rewrite \cref{eq:Psi-same} as
	\begin{equation}
	\label{eq:Psi-same-compl}
	\begin{gathered}
	\ket{\Psi_{\text{same}}} =  \frac{  \left(\ket{++}+\ket{--}\right)_{AB} \hat{M}_{ab} \ket{\lambda_{\rm even}}_{ab} + \frac{1}{2} \left(\ket{+-}+\ket{-+}\right)_{AB} \hat{M}_{ab} \ket{\lambda_{\rm odd}}_{ab}}{2 \sqrt{p_{\rm succ,same}}},
	\end{gathered}
	\end{equation}
	with $\ket{\lambda_{\text{even}}}_{ab}$ and $\ket{\lambda_{\text{odd}}}_{ab}$ being unnormalised states defined as
	\begin{gather}
	\label{eq:lambda_even}
	\ket{\lambda_{\text{even}}}_{ab} = \frac{1}{2} (\ket{\sqrt{\mu}}_a\ket{\sqrt{\mu}}_b+\ket{-\sqrt{\mu}}_a\ket{-\sqrt{\mu}}_b) = \sum_{n \in \mathds{N}_0}  \sqrt{P_{n\vert \mu}} \ket{\lambda_{n}}_{ab}, \\ 	\label{eq:lambda_odd}
	\ket{\lambda_{\text{odd}}}_{ab} = \frac{1}{2} (\ket{\sqrt{\mu}}_a\ket{\sqrt{\mu}}_b-\ket{-\sqrt{\mu}}_a\ket{-\sqrt{\mu}}_b) = \sum_{n \in \mathds{N}_1}  \sqrt{P_{n\vert \mu}} \ket{\lambda_{n}}_{ab}, 			
	\end{gather}
	where $\mathds{N}_0$ ($\mathds{N}_1$) is the set of non-negative even (odd) numbers, $\ket{\lambda_{n}}$ is the $n$-photon two-mode Fock state defined by
	\begin{equation}
	\ket{\lambda_{n}}_{ab} = \frac{1}{\sqrt{2^n n!}} (a^\dagger+b^\dagger)^n \ket{00}_{ab},
	\end{equation}
	and
	\begin{equation}
	\label{eq:P_n_mu}
	\begin{gathered}
	P_{n \vert \mu} = \frac{e^{-2 \mu} (2\mu)^{n}}{n!}, \\
	\end{gathered}
	\end{equation}
	follows a Poisson distribution of average $2\mu$. Combining \cref{eq:ph-error-same} and \cref{eq:Psi-same-compl}, we have that
	\begin{equation}
	\label{eq:ph-error-same2}
	e_{\text{ph,same}} = \frac{1}{2 p_{\rm succ,same}} \norm{\hat{M}_{ab} \ket{\lambda_{\rm even}}_{ab} }^2.
	\end{equation}
	How to estimate the quantity in \cref{eq:ph-error-same2} would be critical in our security proof. One possible approach would be to apply the Cauchy-Schwarz inequality to show that
	\begin{equation}
	\label{eq:cauchy-schwarz}
	\norm{\hat{M}_{ab} \ket{\lambda_{\rm even}}_{ab} }^2  \leq \left[\sum_{n \in \mathds{N}_0} \sqrt{P_{n \vert \mu} Y_n} \right]^2,
	\end{equation}
	where $Y_n = \norm{\hat{M}_{ab}\ket{\lambda_n}_{ab}}^2$ is the yield probability of the state $\ket{\lambda_n}_{ab}$. Let us assume that Alice and Bob used  continuous phase-randomisation on their test mode emissions, and kept only the data from the events in which they use the same intensity and the same phase. Then, the resulting post-selected state, given that they both chose intensity $\beta$, can be expressed as
	\begin{equation}
	\label{eq:beta-continuous-phase-rand}
	\frac{1}{2 \pi} \int_0^{2 \pi} d\theta \ket{\sqrt{\beta} e^{i \theta}} \ket{\sqrt{\beta} e^{i \theta}} \! \bra{\sqrt{\beta} e^{i \theta}} \bra{\sqrt{\beta} e^{i \theta}}_{ab} = \sum_{n=0}^{\infty} P_{n \vert \beta} \ketbra{\lambda_{n}}_{ab},
	\end{equation}
	where $P_{n \vert \beta}$ follows a Poisson distribution and is given by \cref{eq:P_n_mu}. Then, one could apply the standard decoy state method to estimate the yield probabilities $Y_n$, $\forall n \in \mathds{N}_0$, and plug these in \cref{eq:cauchy-schwarz} to estimate \cref{eq:ph-error-same2}. This approach is similar to that of Ref.~\cite{wang2019twinfield}. However, note that if Alice and Bob used continuous phase-randomisation, the probability that Alice and Bob select exactly the same phase $\theta$ is zero, and the resulting protocol is not implementable in practice.
	
	Here, we use the same test-mode phase-postselection idea as in Ref.~\cite{wang2019twinfield}, but we employ discrete phase-randomisation, which results in a protocol that is actually implementable in practice. In this case, \cref{eq:beta-continuous-phase-rand} becomes
	\begin{equation}
	\label{eq:beta-state}
	\rho_{\beta} = \frac{1}{M} \sum_{m = 0}^{M-1} \ket{\sqrt{\beta} e^{\frac{2 i \pi m}{M}}} \ket{\sqrt{\beta} e^{\frac{2 i \pi m}{M}}} \! \bra{\sqrt{\beta} e^{\frac{2 i \pi m}{M}}} \bra{\sqrt{\beta} e^{\frac{2 i \pi m}{M}}}_{ab} = \sum_{n=0}^{M-1} P_{n\,\text{mod}\,M}^{\beta} \ketbra{\lambda_{n\,\text{mod}\,M}^{\beta}}_{ab},
	\end{equation}
	where $\rho_{\beta}$ is the post-selected state when Alice and Bob both used intensity $\beta$ and the same phase \cite{cao2015discrete}. In \cref{eq:beta-state}, we have
	\begin{gather}
	\ket{\lambda_{n\,\text{mod}\,M}^{\beta}}_{ab} =  \sum_{l=0}^{\infty} \sqrt{\frac{P_{Ml+n \vert \beta}}{P_{n\,\text{mod}\,M}^{\beta}}} \ket{\lambda_{Ml+n}}_{ab}, \\ \label{eq:P_nmodM_beta}
	P_{n\,\text{mod}\,M}^{\beta} = \sum_{l=0}^{\infty} P_{Ml+n \vert \beta}.
	\end{gather}
	and $P_{n\vert\beta}$ is given by \cref{eq:P_n_mu}.
	
	Note that for the vacuum intensity $\beta_v$, we have that
	\begin{equation}
	\label{eq:rhobetav}
	\rho_{\beta_v} = \ketbra{\lambda_{0\,\text{mod}\,M}^{\beta_v}}_{ab} = \ketbra{\lambda_0}_{ab}.
	\end{equation}

	Unlike the states $\ket{\lambda_n}$ in \cref{eq:beta-continuous-phase-rand}, the states $\ket{\lambda_{n\,\text{mod}\,M}^{\beta}}$ in \cref{eq:beta-state} have a slight dependence on the intensity $\beta$.  Thus, their yield probabilities,
	\begin{equation}
	\label{eq:yield}
	Y_{n\,\text{mod}\,M}^{\beta} = \norm{\hat{M}_{ab} \ket{\lambda_{n\,\text{mod}\,M}^{\beta}}_{ab}}^2,
	\end{equation}
	are not equal for two different intensities $\beta_1$ and $\beta_2$, which prevents us from applying the standard decoy-state method. Instead, we use a similar idea as in Ref.~\cite{primaatmaja2019versatile}, defining the Gram matrix $G$ of the set of Eve's post-measurement states, and constructing a semidefinite program where the objective function and all the constraints are linear functions of entries of $G$. In our case, we define $G$ as the Gram matrix of the vector set $\left\{\hat{M}_{ab}\ket{ \lambda_{n\,\text{mod}\,M}^{\beta}}\right\}$, $\forall \beta \in \mathcal{T}$ and $n \in \{0,1,...,M-1\}$, where $\mathcal{T}$ is the set of all test-mode intensities, except vacuum. The entries of $G$ are $G_{ij} = \braket{i}{j}$, where $\ket{i}$ denotes the $i$-th element of the vector set.

	Our objective function is \cref{eq:ph-error-same2}, which we can write as 
	\begin{equation}
	\label{eq:ph-error-same3}
	e_{\text{ph,same}} = \frac{1}{2 p_{\rm succ,same}} \mel{\lambda_{\rm even}}{\hat{M}_{ab}^{\dagger} \hat{M}_{ab}}{\lambda_{\rm even}}
	\end{equation}
	By re-expressing $\ket{\lambda_{\rm even}}$ and  $\ket{\lambda_{\rm odd}}$ in \cref{eq:lambda_even,eq:lambda_odd} as
	\begin{equation}
	\begin{gathered}
	\ket{\lambda_{\text{even}}}_{ab} = \sum_{\substack{n=0 \\ n\in \mathds{N}_0}}^{M-1} \sqrt{P_{n\,\text{mod}\,M}^{\mu}} \ket{\lambda_{n\,\text{mod}\,M}^{\mu}}_{ab}, \\
	\ket{\lambda_{\text{odd}}}_{ab}  = \sum_{\substack{n=0 \\ n\in \mathds{N}_1}}^{M-1} \sqrt{P_{n\,\text{mod}\,M}^{\mu}} \ket{\lambda_{n\,\text{mod}\,M}^{\mu}}_{ab},
	\end{gathered}
	\end{equation}
	it becomes clear that the right-hand side of \cref{eq:ph-error-same3} is a linear function of elements of $G$.
	
	We obtain our first constraint by taking the norm squared of both sides of \cref{eq:Psi-same}, and solving for $p_{\rm succ,same}$, obtaining
	\begin{equation}
	p_{\rm succ,same} = \frac{1}{2} \mel{\lambda_{\rm even}}{\hat{M}^{\dagger} \hat{M}}{\lambda_{\rm even}} + \frac{1}{2} \mel{\lambda_{\rm odd}}{\hat{M}^{\dagger} \hat{M}}{\lambda_{\rm odd}}.
	\end{equation}
	From \cref{eq:beta-state}, we have that
	\begin{equation}
	\label{eq:Qbeta}
	Q_{\beta} = \sum_{n=0}^{M-1} P_{n\,\text{mod}\,M}^{\beta} Y_{n\,\text{mod}\,M}^{\beta},
	\end{equation}
	where $Q_{\beta}$ is the measured gain of the state $\rho_{\beta}$. Note that $Y_{n\,\text{mod}\,M}^{\beta}$ is a (diagonal) element of $G$, thus \cref{eq:Qbeta} is a linear function of elements of $G$.
	
	The next constraints use the trace distance inequality \blue{\cite{cao2015discrete}}
	\begin{equation}
	\label{eq:trace-distance-bound}
	\begin{gathered}
	Y_{n\,\text{mod}\,M}^{\beta_1} -	Y_{n\,\text{mod}\,M}^{\beta_2} \leq \sqrt{1- F_{n}^{\beta_1,\beta_2}}, \\
	\end{gathered}
	\end{equation}
	where
	\begin{equation}
	\label{eq:F_n_beta1_beta2}
	F_{n}^{\beta_1,\beta_2} = \abs{\braket{\lambda_{n\,\text{mod}\,M}^{\beta_1}}{\lambda_{n\,\text{mod}\,M}^{\beta_2}}_{ab}}^2 = \left[\sum_{l=0}^{\infty} \sqrt{\frac{P_{Ml+n \vert \beta_1}}{P_{n\,\text{mod}\,M}^{\beta_1}}}  \sqrt{\frac{P_{Ml+n \vert \beta_2}}{P_{n\,\text{mod}\,M}^{\beta_2}}}\right]^2.
	\end{equation}
	Our next constraint is based on the inequality
	\begin{equation}
	\label{eq:lo-preskill-bound}
	Y_{n\,\text{mod}\,M}^{\beta_1} \leq 1- Y_{n\,\text{mod}\,M}^{\beta_2} + 2 \sqrt{F_{n}^{\beta_1,\beta_2}(1-F_{n}^{\beta_1,\beta_2}) (1-Y_{n\,\text{mod}\,M}^{\beta_2}) Y_{n\,\text{mod}\,M}^{\beta_2}} +F_{n}^{\beta_1,\beta_2} (2Y_{n\,\text{mod}\,M}^{\beta_2}-1),
	\end{equation}
	which is tighter than the trace distance inequality in \cref{eq:trace-distance-bound}, but cannot be added to the SDP, since it is a non-linear function of $Y_{n\,\text{mod}\,M}^{\beta_2}$, an element of $G$. The only exception is the case $n=0$ and $\beta_2 = \beta_v$, since from \cref{eq:rhobetav}, we have that
	\begin{equation}
	Y_{0\,\text{mod}\,M}^{\beta_v} = Y_0 = Q_{\beta_v},
	\end{equation}
	and $Q_{\beta_v}$ is directly measurable from the protocol.
	Thus, substituting $n=0$, $\beta_1 = \beta$, $\beta_2 = \beta_v$ and 	$Y_{0\,\text{mod}\,M}^{\beta_v} =  Q_{\beta_v}$ in \cref{eq:lo-preskill-bound}, we have the inequality
	\begin{equation}
	Y_{0\,\text{mod}\,M}^{\beta} \leq 1- Q_{\beta_v} + 2 \sqrt{F_{0}^{\beta,\beta_v}(1-F_{0}^{\beta,\beta_v}) (1-Q_{\beta_v}) Q_{\beta_v}} +F_{0}^{\beta,\beta_v} (2 Q_{\beta_v}-1),
	\end{equation}
	which is a linear function of $Y_{0\,\text{mod}\,M}^{\beta}$.
	
	For our final constraints, we use the fact that $Y_{n\,\text{mod}\,M}^{\beta} \leq 1$, $\forall n, \beta$. To reduce the number of constraints, we only include the case $\beta = \mu$.
	
	Combining everything, we have that our upper-bound on $e_{\text{ph,same}}$ is the solution of the following SDP
	\begin{equation}
	\label{eq:SDP}
	\begin{aligned}
	&\max_{\blue{G}} ~~  \frac{1}{2 p_{\rm succ,same}} \mel{\lambda_{\rm even}}{\hat{M}^{\dagger} \hat{M}}{\lambda_{\rm even}} \text{  s.t. } \\
	& p_{\rm succ,same} = \frac{1}{2} \mel{\lambda_{\rm even}}{\hat{M}^{\dagger} \hat{M}}{\lambda_{\rm even}} + \frac{1}{2} \mel{\lambda_{\rm odd}}{\hat{M}^{\dagger} \hat{M}}{\lambda_{\rm odd}}, \\
	& Q_{\beta} = \sum_{n=0}^{M-1} P_{n\,\text{mod}\,M}^{\beta} Y_{n\,\text{mod}\,M}^{\beta}, ~~ \forall \beta \in \mathcal{T},
	\\
	& Y_{n\,\text{mod}\,M}^{\mu} \leq 1, ~~ \forall n \in \{0,...,M-1\}, \\
	&  Y_{n\,\text{mod}\,M}^{\beta_1} - Y_{n\,\text{mod}\,M}^{\beta_2} \leq \sqrt{1- F_{n}^{\beta_1,\beta_2}}, ~~ \forall \beta_1,\beta_2 \in \mathcal{T},~n \in \{0,...,M-1\},\\
	&Y_{0\,\text{mod}\,M}^{\beta} \leq 1- Q_{\beta_v} + 2 \sqrt{F_{0}^{\beta,\beta_v}(1-F_{0}^{\beta,\beta_v}) (1-Q_{\beta_v}) Q_{\beta_v}} +F_{0}^{\beta,\beta_v} (2 Q_{\beta_v}-1), ~~ \forall \beta \in \mathcal{T},
	\end{aligned}
	\end{equation}
	where $\mathcal{T} = \{\beta_1,\ldots, \beta_{d-2}, \mu\}$ is the set of all test-mode intensities, except vacuum.

	\subsubsection{Estimation of \texorpdfstring{$e_{\rm ph,diff}$}{ephdiff}}
	
	The procedure to estimate $e_{\rm ph,diff}$ is very similar to that of $e_{\rm ph,same}$. In this case, we rewrite \cref{eq:Psi-diff} as
	\begin{equation}
	\label{eq:Psi-diff-compl}
	\begin{gathered}
	\ket{\Psi_{\text{diff}}} =  \frac{  \left(\ket{++}\blue{-}\ket{--}\right)_{AB} \hat{M}_{ab} \ket{\lambda_{\rm even}^{-}}_{ab} + \frac{1}{2} \left(\ket{-+}\blue{-}\ket{+-}\right)_{AB} \hat{M}_{ab} \ket{\lambda_{\rm odd}^{-}}_{ab}}{2 \sqrt{p_{\rm succ,diff}}},
	\end{gathered}
	\end{equation}
	where $\ket{\lambda_{\rm even}^{-}}$ and $\ket{\lambda_{\text{odd}}^{-}}$ are unnormalised states defined as
	\begin{gather}
	\ket{\lambda_{\text{even}}^{-}}_{ab} = \frac{1}{2} (\ket{\sqrt{\mu}}_a\ket{-\sqrt{\mu}}_b+\ket{-\sqrt{\mu}}_a\ket{\sqrt{\mu}}_b) = \sum_{n+m \in \mathds{N}_0} c_n c_m \ket{n}_a \ket{m}_b = \sum_{n \in \mathds{N}_0}  \sqrt{P_{n\vert \mu}} \ket{\lambda_{n}^{-}}_{ab}, \\
	\ket{\lambda_{\text{odd}}^{-}}_{ab} = \frac{1}{2} (\ket{\sqrt{\mu}}_a\ket{-\sqrt{\mu}}_b-\ket{-\sqrt{\mu}}_a\ket{\sqrt{\mu}}_b) = \sum_{n+m \in \mathds{N}_1} c_n c_m \ket{n}_a \ket{m}_b = \sum_{n \in \mathds{N}_1}  \sqrt{P_{n\vert \mu}} \ket{\lambda_{n}^{-}}_{ab}, 			
	\end{gather}
	$\ket{\lambda_{n}^{-}}$ is the $n$-photon two-mode Fock state defined by
	\begin{equation}
	\ket{\lambda_{n}^{-}}_{ab} = \frac{1}{\sqrt{2^n n!}} (a^\dagger-b^\dagger)^n \ket{00}_{ab},
	\end{equation}
	and $P_{n\vert\mu}$ is given by \cref*{eq:P_n_mu}. In this case, the state after post-selecting the test mode emissions in which Alice and Bob both used intensity $\beta$ and the opposite phases $\theta_a = \theta_b \pm \pi = \theta$ is
	\begin{equation}
	\label{eq:beta-state-diff}
	\rho_{\beta}^{-} = \frac{1}{M} \sum_{m = 0}^{M-1} \ket{\sqrt{\beta} e^{\frac{2 i \pi m}{M}}} \ket{-\sqrt{\beta}e^{\frac{2 i \pi m}{M}}} \bra{\sqrt{\beta}e^{\frac{2 i \pi m}{M}}} \bra{-\sqrt{\beta} e^{\frac{2 i \pi m}{M}}}_{ab} = \sum_{n=0}^{M-1} P_{n\,\text{mod}\,M}^{\beta} \ketbra{\lambda_{n\,\text{mod}\,M}^{\beta,-}}_{ab},
	\end{equation}
	where
	\begin{gather}
	\ket{\lambda_{n\,\text{mod}\,M}^{\beta,-}}_{ab} =  \sum_{l=0}^{\infty} \sqrt{\frac{P_{Ml+n \vert \beta}}{P_{n\,\text{mod}\,M}^{\beta}}} \ket{\lambda_{Ml+n}^{-}}_{ab},
	\end{gather}
	$P_{n\vert\beta}$ is given by \cref{eq:P_n_mu}, and  $P_{n\,\text{mod}\,M}^{\beta}$ is given by \cref{eq:P_nmodM_beta}. 
	
	Similarly as in the previous subsection, we re-express $\ket{\lambda_{\rm even}^{-}}$ and  $\ket{\lambda_{\rm odd}^{-}}$ as
	\begin{equation}
	\begin{gathered}
	\ket{\lambda_{\text{even}}^{-}}_{ab} = \sum_{\substack{n=0 \\ n\in \mathds{N}_0}}^{M-1} \sqrt{P_{n\,\text{mod}\,M}^{\mu}} \ket{\lambda_{n\,\text{mod}\,M}^{\mu,-}}_{ab}, \\
	\ket{\lambda_{\text{odd}}^{-}}_{ab}  = \sum_{\substack{n=0 \\ n\in \mathds{N}_1}}^{M-1} \sqrt{P_{n\,\text{mod}\,M}^{\mu}} \ket{\lambda_{n\,\text{mod}\,M}^{\mu,-}}_{ab},
	\end{gathered}
	\end{equation}
	and define
	\begin{equation}
	\label{eq:yield_minus}
	Y_{n\,\text{mod}\,M}^{\beta,-} = \norm{\hat{M}_{ab} \ket{\lambda_{n\,\text{mod}\,M}^{\beta,-}}_{ab}}^2.
	\end{equation}
	
	This time, we define $G$ as the Gram matrix of the vector set $\left\{\hat{M}_{ab}\ket{ \lambda_{n\,\text{mod}\,M}^{\beta,-}}\right\}$, and follow a similar procedure as in last subsection to construct the objective function and the constraints. In the end, we have that our upper-bound on $e_{\text{ph,diff}}$ is the solution of the following SDP
	\begin{equation}
	\label{eq:SDP_minus}
	\begin{aligned}
	&\max_{\blue{G}} ~~  \frac{1}{2 p_{\rm succ,diff}} \ev{\hat{M}^{\dagger} \hat{M}}{\lambda_{\rm even}^{-}} \text{  s.t. } \\
	& p_{\rm succ,diff} = \frac{1}{2} \ev{\hat{M}^{\dagger} \hat{M}}{\lambda_{\rm even}^{-}} + \frac{1}{2} \ev{\hat{M}^{\dagger} \hat{M}}{\lambda_{\rm odd}^{-}}, \\
	& Q_{\beta}^{-} = \sum_{n=0}^{M-1} P_{n\,\text{mod}\,M}^{\beta} Y_{n\,\text{mod}\,M}^{\beta,-}, ~~ \forall \beta \in \mathcal{T},
	\\
	& Y_{n\,\text{mod}\,M}^{\mu,-} \leq 1, ~~ \forall n \in \{0,...,M-1\}, \\
	&  Y_{n\,\text{mod}\,M}^{\beta_1,-} - Y_{n\,\text{mod}\,M}^{\beta_2,-} \leq \sqrt{1- F_{n}^{\beta_1,\beta_2}}, ~~ \forall \beta_1,\beta_2 \in \mathcal{T},~n \in \{0,...,M-1\},\\
	&Y_{0\,\text{mod}\,M}^{\beta,-} \leq 1- Q_{\beta_v}^{-} + 2 \sqrt{F_{0}^{\beta,\beta_v}(1-F_{0}^{\beta,\beta_v}) (1-Q_{\beta_v}^{-} )Q_{\beta_v}^{-} } +F_{0}^{\beta,\beta_v} (2 Q_{\beta_v}^{-} -1), ~~ \forall \beta \in \mathcal{T},
	\end{aligned}
	\end{equation}
	where $F_{n}^{\beta_1,\beta_2}$ is given by \cref{eq:F_n_beta1_beta2} and  $\mathcal{T} = \{\beta_1,\ldots, \beta_{d-2}, \mu\}$ is the set of all test-mode intensities, except vacuum.

	\section{Numerical results}
	
	Here, we simulate the secret key rate obtainable as a function of the overall Alice-Bob loss, which includes the inefficiency of Charlie's detectors, for different values of $M$, the number of random phases. For the sake of our numerical simulations, we assume that there is no eavesdropper, and we only model the imperfections in the system to simulate the values one may obtain in a real experiment. We assume a misalignment error rate of $2\%$, matching the results of a recent experiment \cite{minder2019experimental}, and a dark count probability of $10^{-8}$ per pulse. In all curves, we assume that Alice and Bob use three different test-mode intensities $\{\beta_1, \mu, \beta_v\}$, where $\beta_v = 0$ is a vacuum intensity and $\mu$ is the same intensity used in key mode. We optimise over the value of $\mu$ and $\beta_1$, with the condition that $\mu,\beta_1 \geq 10^{-4}$. \blue{This condition is motivated by the fact that it is experimentally difficult to produce a laser pulse with a very small, but fixed, intensity.} 
	
\begin{figure}[ht]
		\centering		\resizebox{.75\linewidth}{!}{\input{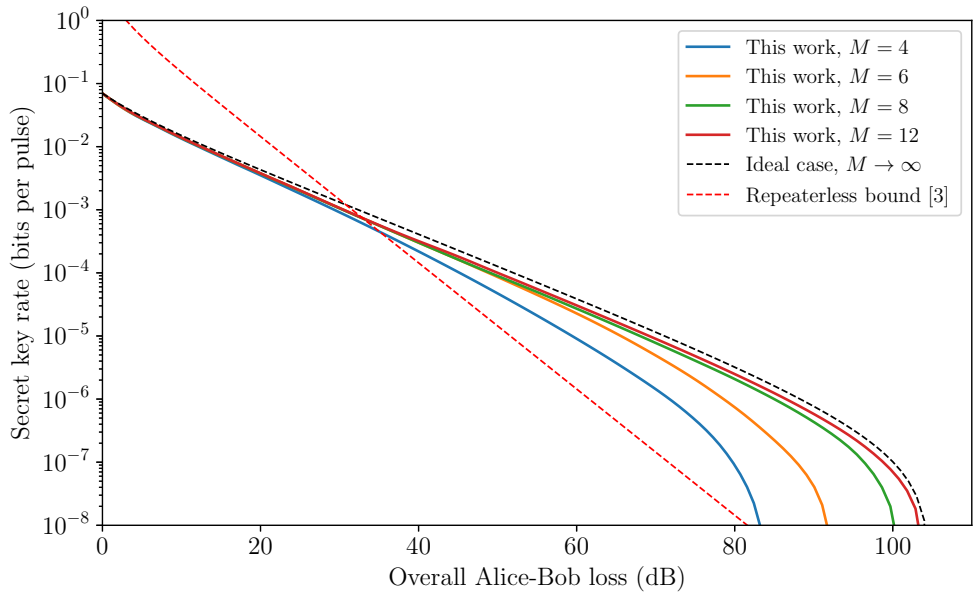}}
		\caption{Secret key rate for our discrete phase protocol at different values of $M$, in comparison to fundamental bound for repeaterless QKD systems $-\log_2(1-\eta)$, where $\eta$ is the overall Alice-Bob transmissivity.}
		\label{fig:graph1}
	\end{figure}

	In \cref{fig:graph1}, we see that the protocol can overcome the repeaterless bound \cite{pirandola2017fundamental} with as few as four random phases. For the ideal case of $M \to \infty$, we use \cref{eq:cauchy-schwarz}, assuming that Alice and Bob are somehow able to estimate the exact values of $Y_n$, $\forall n$, using the data collected in test mode.  As explained in the discussion following \cref{eq:cauchy-schwarz}, the case of $M \to \infty$ is not actually implementable in practice, but it provides an upper-bound on the secret key rate obtainable \blue{for finite values of $M$}. Notably, \cref{fig:graph1} shows that one can get very close to this ideal scenario with only $M=12$ random phases.

\begin{figure}[ht]
		\centering		\resizebox{.75\linewidth}{!}{\input{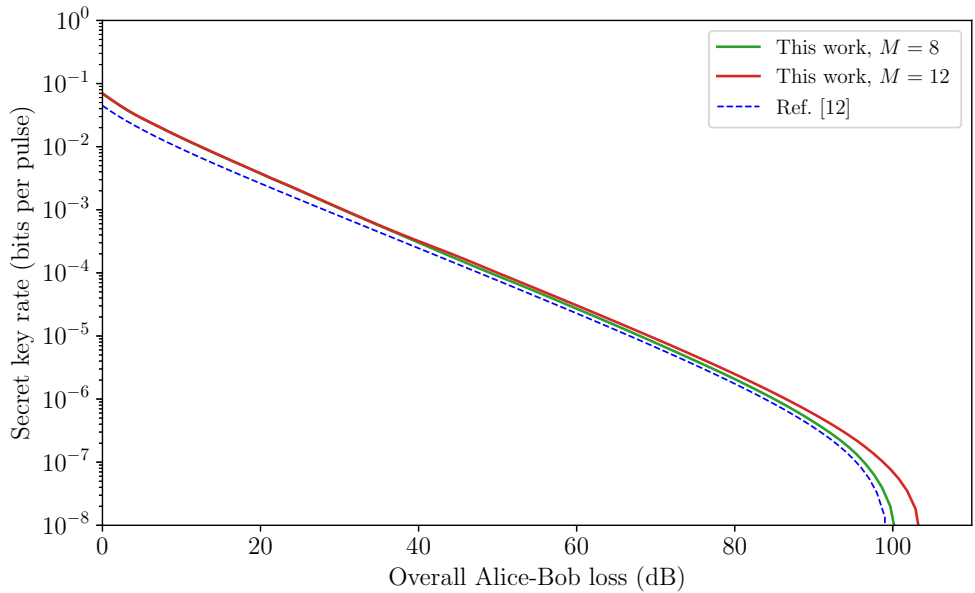}}
		\caption{Comparison between the results of this work and those of Ref.~\cite{curty2019simple}, which uses continuous phase randomisation in its test-mode emissions. For simplicity, to compute the results in \cite{curty2019simple}, we assume that Alice and Bob's test-mode rounds provide perfect estimates of the yield probabilities $Y_{nm}$ for $n+m \leq 4$, while the rest are upper-bounded by one. This is an ideal scenario and, as shown in \cite{curty2019simple}, the results will be slightly worse once one considers the imperfect estimates that result from the use of a finite set of decoy states, as we do for the results in this work.}
		\label{fig:graphcomp}
	\end{figure}

\begin{figure}[ht]
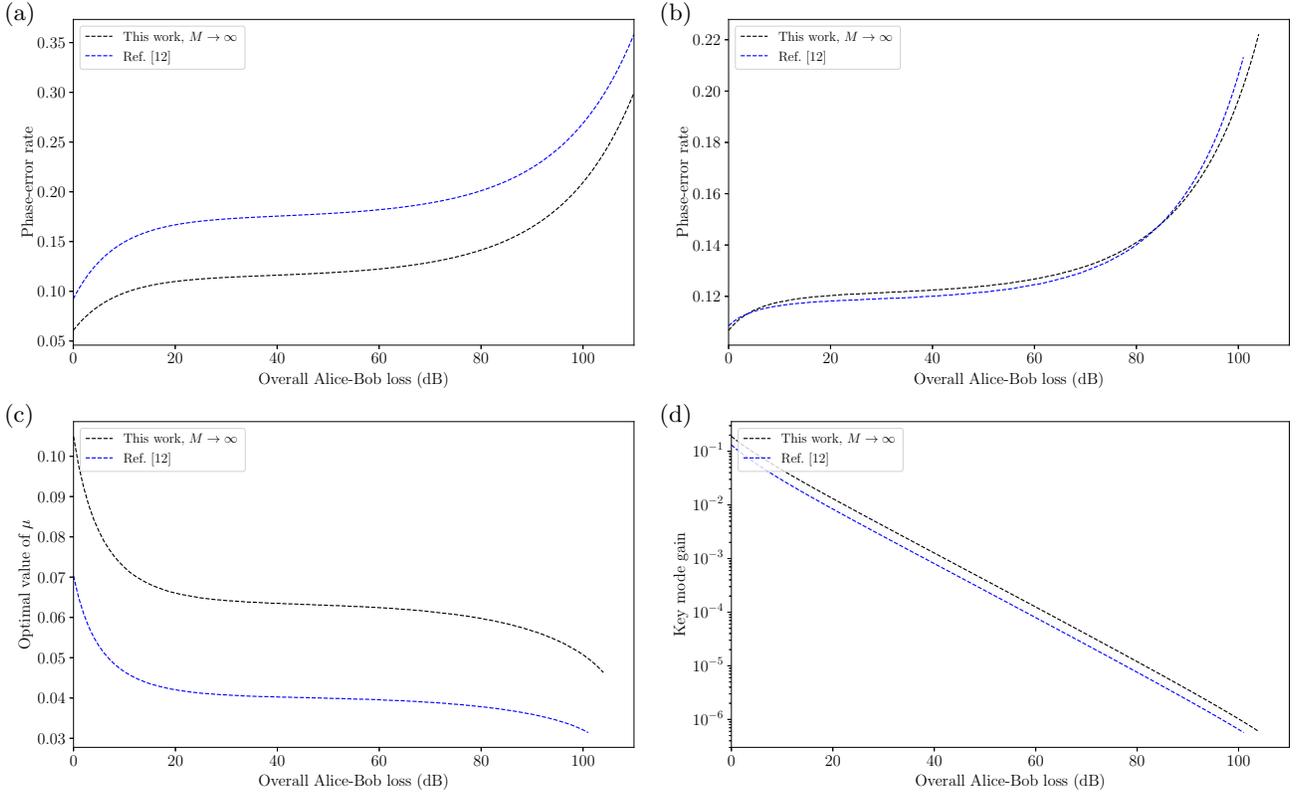

		\centering

		\topinset{(a)}{\resizebox{.48\linewidth}{!}{\input{figs/extra/graph_phase_error_fixed.pgf}}}{0in}{0in}
		 \topinset{(b)}{\resizebox{.48\linewidth}{!}{\input{figs/extra/graph_phase_error_opt_each.pgf}}}{0in}{0in}
		 \topinset{(c)}{\resizebox{.48\linewidth}{!}{\input{figs/extra/graph_opt_mu.pgf}}}{0in}{0in}
		 \topinset{(d)}{\resizebox{.48\linewidth}{!}{\input{figs/extra/graph_opt_gain.pgf}}}{0in}{0in}
    		\caption{Comparison between the value of some terms in our analysis, for the ideal case $M \to \infty$, and the analysis in Ref.~\cite{curty2019simple}. (a) Upper-bound on the phase-error rate, assuming a fixed value $\mu = 0.06$. (b) Upper-bound on the phase-error rate, for the value of $\mu$ that optimises the key rate in each analysis. (c) Value of $\mu$ that optimises the key rate in each analysis. (d) Key mode gain for the value of $\mu$ that optimises the key rate in each analysis. }
		\label{fig:graphextra}
	\end{figure}

	In \cref{fig:graphcomp}, we compare the results of our protocol with those of Ref.~\cite{curty2019simple}. The latter is one of the best performing variants of TF-QKD, in both the asymptotic \cite{LucamariniSlides} and finite-key \cite{curras2019tight} regimes, that employ continuous phase randomisation. Moreover, its quantum phase is very close to ours, with the only difference being their use of continuous phase randomisation in test mode. Thus, \cref{fig:graphcomp} directly compares the performance of the discrete and continuous randomisation approaches. Remarkably, we obtain higher secret-key rates using discrete phase randomisation, as long as one uses eight random phases or more. This may sound surprising, at first instance, but it is justified by the fact that, for the same value of $\mu$, we can obtain a tighter estimation of the phase-error rate in the discrete-phase version, thanks to the test-mode phase postselection. This can be seen in \cref{fig:graphextra}(a), where we compare the upper-bound on the phase-error rate of the two protocols for a fixed value $\mu = 0.06$. In a practical setting, one would optimise over the value of $\mu$, in which case the two protocols result in similar bounds for the phase-error rate, see \cref{fig:graphextra}(b). But, this will be achieved at a higher value of $\mu$ for our protocol, see \cref{fig:graphextra}(c), which results in a higher gain, see \cref{fig:graphextra}(d), hence a higher secret-key rate.

	\section{Conclusion and discussion}
	
	Most previous variants of TF-QKD have relied on the emission of weak laser pulses with a continuous random phase, which is difficult to achieve and certify in practice. Here, we  proposed a practical TF-QKD variant that used discrete phase randomisation instead. Its security proof relied on post-selecting the test-mode rounds in which both users employed exactly matching phase values, which is not practically possible with a continuous randomisation approach. We consequently obtained \textit{higher} key rates using discrete randomisation. This is interesting, given that discrete randomisation is usually considered to be a source flaw. In fact, previous analyses of decoy-state QKD with discrete randomisation \cite{cao2015discrete} obtained strictly worse results than their continuous counterparts. Our security analysis relied on a customised version of numerical techniques for MDI-QKD protocols based on semidefinite programming, which had a substantially reduced complexity as compared with the generic approach.
	
	In our analysis, we have considered the asymptotic regime in which Alice and Bob run the protocol for infinitely many rounds. It remains an open question whether discrete randomisation could still offer an advantage in a finite-key setting. Since state-of-the-art numerical finite-key proofs can only prove security tightly against a restricted class of eavesdropping attacks \cite{bunandar2019numerical,george2020numerical}, important developments are needed before we can rigorously answer this question.
	
	\section*{Acknowledgements}
	
	We thank Kiyoshi Tamaki, Marcos Curty, Álvaro Navarrete, Margarida Pereira, Zhen-Qiang Yin, and Xiongfeng Ma for valuable discussions. This work was supported by the European Union's Horizon 2020 research and innovation programme under the Marie Sklodowska-Curie grant agreement number 675662 (QCALL). All data generated can be reproduced by the equations and the methodology introduced in this paper.

	\bibliography{refs}

\end{document}